%% file: paper.tex
\documentstyle[aps,psfig,multicol]{revtex}

\begin{document}

\draft

\title{Nucleon-Nucleon Phase Shifts and Pairing in 
Neutron Matter and Nuclear Matter}
 
\author{\O.\ Elgar\o y$^a$ and M.\ Hjorth-Jensen$^b$}

\address{$^a$Department of Physics, University of Oslo, N-0316 Oslo, Norway}

\address{$^b$Nordita, Blegdamsvej 17, DK-2100 K\o benhavn \O, Denmark}

\maketitle

\begin{abstract}

We consider $^1S_0$ pairing in infinite neutron matter 
and nuclear matter 
and show that in the lowest order approximation, where the pairing 
interaction is taken to be the bare nucleon-nucleon (NN) interaction 
in the $^1S_0$ channel, the pairing interaction and the energy gap 
can be determined directly from the $^1S_0$ phase shifts.  
This is due to the almost separable 
character of the nucleon-nucleon interaction in this partial wave. 
Since the most recent NN interactions are charge-dependent, we 
have to solve coupled gap equations for proton-proton, neutron-neutron, 
and neutron-proton pairing in nuclear matter.  The results are, 
however, found to be close to those obtained with charge-independent 
potentials.

\end{abstract}


\begin{multicols}{2}

Recently, there has been renewed interest in the pairing problem in 
neutron matter and neutron-rich nuclei.  The superfluid properties 
of neutron matter is of importance in the study of neutron stars 
\cite{petra95}, while pairing in neutron-rich systems is of relevance 
for the study of heavy nuclei close to the drip line \cite{mulshe93} 
and the light halo nuclei \cite{riis94}.  Much effort has gone into 
calculating the superfluid energy gap in dilute neutron matter 
\cite{baldo90,chen93,tak93,elg96,khodel96}.  
Most of these studies,  
e.g., those of Refs.\ \cite{baldo90,tak93,elg96,khodel96} have been carried 
out using pairing matrix elements given by the bare nucleon-nucleon 
(NN) interaction.  Many of the same authors have calculated the 
$^1S_0$ gap in nuclear matter, which has also been the subject of  
recent relativistic formulations of the pairing problem 
\cite{ring90,guim96,matera97}.  

Even though it is a long time since  
Clark et al.\ \cite{clark76} showed that the effects of density and 
spin-density fluctuations must be included in the pairing interaction, 
and there 
has been much progress in that direction recently \cite{wam93,schulze96}, 
we will here focus on the situation at the level of the 
bare interaction.  In this lowest-order approximation to the 
problem it has been found that results for the $^1S_0$ energy 
gap in neutron matter and in nuclear matter are almost independent 
of the choice of NN interaction.  
We aim at explaining how this can 
be understood directly from the measured properties of the free NN 
interaction.  
Moreover, although a relation between the pairing gap and
NN phase shifts was obtained almost forty years ago by Emery and
Sessler \cite{es60} (see also Hoffberg et al.\ \cite{hoffberg70}),  
in this work we wish to focus on the near 
interaction independence of the results for the energy gap at the 
Fermi level, and try to explain this  
from the NN scattering data directly. 

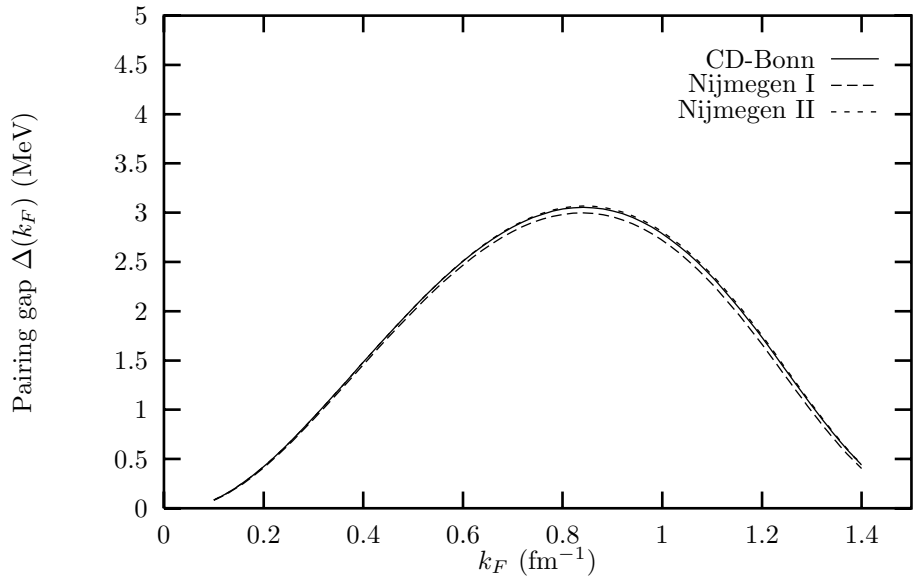
\begin{figure}
    \input{fig1.tex}
    \caption{$^1S_0$ energy gap in neutron matter with the CD-Bonn, 
             Nijmegen I and Nijmegen II potentials.}
    \label{fig:fig1}
\end{figure}

The energy gap in infinite matter is obtained by solving the BCS equation 
for the gap function $\Delta(k)$.  
\begin{equation}
      \Delta(k)=-\frac{1}{\pi}\int_{0}^{\infty}dk'k'^2 
                 V(k,k')\frac{\Delta(k')}{E(k')}, 
      \label{eq:eq1}
\end{equation}
where $V(k,k')$ is the bare momentum-space NN interaction in the 
$^1S_0$ channel, and $E(k)$ is the quasiparticle energy given by 
$E(k)=\sqrt{(\epsilon(k)-\epsilon(k_F))^2+\Delta(k)^2}$, where 
$\epsilon(k)$ is the single-particle energy of a neutron with 
momentum $k$, and $k_F$ is the Fermi momentum.  
Medium effects should 
be included in $\epsilon(k)$, but we will use free single-particle 
energies $\epsilon(k)=k^{2}/2m$, where $m$ is the neutron rest mass,  
to avoid unnecessary complications.  
And in neutron matter, at least at the densities 
considered here,  Brueckner-type calculations \cite{elg96} 
indicate that in-medium single-particle energies do not 
differ much from the free ones. 
The energy gap is defined as $\Delta_F\equiv \Delta(k_F)$.  
Eq.\ (\ref{eq:eq1}) can be 
solved by various techniques, some of which are described in 
Refs.\ \cite{elg96,khodel96}.  
In Fig.\ \ref{fig:fig1} we show 
the results for $\Delta_F$ obtained with the CD-Bonn potential (full line) 
\cite{mach96},  
the Nijmegen I and Nijmegen II potentials (long-dashed line and 
short-dashed line, respectively) \cite{nijm94}. 
The results are virtually identical, with the maximum value 
of the gap varying from 2.98 MeV for the Nijmegen I potential to 3.05 MeV 
for the Nijmegen II potential.  The same insensitivity of the energy gap 
with respect to 
the choice of NN interaction was found in Refs.\ 
\cite{baldo90,elg96,khodel96}. 
We will now discuss how these results can be understood 
from the properties of the NN interaction in the $^1S_0$ channel.

A characteristic feature of $^1S_0$ NN scattering is the large, negative 
scattering length, indicating the presence of  
a nearly bound state at zero scattering energy.  Near a bound state, 
where the NN $T$-matrix has a pole, it can be written in separable form, 
and this implies that the NN interaction itself to a good approximation is 
rank-one separable near this pole \cite{brown76}.   
Thus at low energies we can write 
\begin{equation}
       V(k,k')=\lambda v(k)v(k'),
       \label{eq:eq2}
\end{equation}
where $\lambda$ is a constant.  Then it is easily seen from 
Eq.\ (\ref{eq:eq1}) that the gap function can be written as $\Delta_F v(k)$, 
where $\Delta_F$ is the energy gap.  Inserting this form of 
$\Delta(k)$ into Eq.\ (\ref{eq:eq1}) one obtains 
\begin{equation}
      1=-\frac{1}{\pi}\int_{0}^{\infty}dk'k'^2\frac{\lambda v^2(k')}{E(k')}, 
      \label{eq:eq3}
\end{equation}
which shows that the energy gap $\Delta_F$ is 
determined by the diagonal elements $\lambda v^2(k)$ of the NN interaction.  
The crucial point is that in scattering theory it can be shown that 
the inverse scattering problem, that is, the determination of a 
two-particle potential from the knowledge of the phase shifts at all 
energies, is exactly, and uniquely, solvable for rank-one 
separable potentials \cite{brown76,chadan92}.  Following the notation 
of Ref.\ \cite{brown76} we have 
\begin{equation}
       \lambda v^2(k)=-\frac{k^2+\kappa_B^2}{k^2}
                       \frac{\sin \delta(k)}{k}e^{-\alpha(k)},
       \label{eq:eq4}
\end{equation}
for an attractive potential with a bound state at energy $E=-\kappa_B^2$. 
In our case $\kappa_B=0$.    
Here $\delta(k)$ is the $^1S_0$ phase shift as a function of momentum $k$, 
while $\alpha(k)$ is given by a principle value integral: 
\begin{equation}
       \alpha(k)=\frac{1}{\pi}{\rm P}\int_{-\infty}^{+\infty}dk'
                 \frac{\delta(k')}{k'-k},
       \label{eq:eq5}
\end{equation}
where the phase shifts are extended to negative momenta through 
$\delta(-k)=-\delta(k)$.  Eqs.\ (\ref{eq:eq4}) and (\ref{eq:eq5}) 
can also be rewritten in terms of the Jost function \cite{chadan92}  
as done in Ref.\ \cite{kk97}.

 From this discussion we see that $\lambda v^2(k)$, and therefore also 
the energy gap $\Delta_F$, is completely determined by the $^1S_0$ 
phase shifts.  However, there are two obvious limitations on the 
practical validity of this statement.  First of all, the separable 
approximation can only be expected to be good at low energies, near the 
pole in the $T$-matrix.  Secondly, we see from Eq.\ (\ref{eq:eq5}) that 
knowledge of the phase shifts $\delta(k)$ at all energies is required.  
This is, of course, impossible, and most phase shift 
analyses stop at a laboratory energy $E_{\rm lab}=350$ MeV.  
Strictly speaking, the rank-one separable  approximation to the 
$^1S_0$ interaction breaks down already where the 
$^1S_0$ phase shift changes sign from positive to negative at 
$E_{\rm lab}\approx 248$ MeV, corresponding to a single-particle momentum 
of $k\approx1.73\;{\rm fm}^{-1}$.  However, at low values of $k_F$, knowledge 
of $v(k)$ up to this value of $k$ may actually be enough to determine 
the value of $\Delta_F$, as the integrand in Eq.\ (\ref{eq:eq5}) is 
strongly peaked around $k_F$.  We therefore found it worthwhile to 
try to calculate the energy gap directly from the $^1S_0$ phase shifts 
using Eqs. (\ref{eq:eq3})-(\ref{eq:eq5}). 
A possible improvement to the rank-one
separable approach for potentials which change sign is discussed by
Kwong and K\"ohler \cite{kk97}. 

The input in our calculation is the $^1S_0$ phase shifts taken from  
the recent Nijmegen phase shift analysis \cite{nijm93}. 
We then evaluated $\lambda v^2(k)$ from Eqs. (\ref{eq:eq4}) and 
(\ref{eq:eq5}), using methods described in Ref.\ \cite{davies91} to 
evaluate the principle value integral in Eq.\ (\ref{eq:eq5}). 
Finally, we evaluated the energy gap $\Delta_F$ for various values 
of $k_F$ by solving Eq.\ (\ref{eq:eq3}).  
Numerically the integral on the right-hand side of this equation 
depended very weakly on the momentum structure of $\Delta(k)$, so 
in our calculations we could take $\Delta(k)\approx \Delta_F$ in 
Eq.\ (\ref{eq:eq3}), and thus it became an algebraic equation 
for the energy gap $\Delta_F$.  
The resulting energy gap is plotted in Fig.\ 
\ref{fig:fig2} (dashed line) together with the gap obtained with the 
CD-Bonn potential (full line). 
As the reader can see, the agreement 
between the direct calculation from the phase shifts and the CD-Bonn 
calculation of $\Delta_F$ is very good, even 
at densities as high as $k_F=1.4\;{\rm fm}^{-1}$.  The energy gap 
is to a great extent determined by the available $^1S_0$ phase shifts.  
In the same figure we also report the results (dot-dashed line) 
obtained using the effective range approximation to the phase shifts: 
\begin{equation}
       k\cot \delta(k)=-\frac{1}{a_0}+\frac{1}{2}r_0 k^2,
       \label{eq:eq6}
\end{equation}
where $a_0=-18.8\pm 0.3$ fm and $r_0=2.75\pm 0.11$ fm are the singlet 
neutron-neutron scattering length and effective range, respectively.  
In this case an analytic expression can be obtained for $\lambda v^2(k)$, as 
shown in Ref.\ \cite{chadan92}:
\begin{equation}
       \lambda v^2(k)=-\frac{1}{\sqrt{k^2+
                       \frac{r_0^2}{4}(k^2+\alpha^2)^2}}
                       \sqrt{\frac{k^2+\beta_2^2}{k^2-\beta_1^2}},
       \label{eq:eq7}
\end{equation}
with $\alpha^2=-2/a_0 r_0$, and $\beta_1\approx-0.0498\;{\rm fm}^{-1}$, 
and $\beta_2\approx 0.777\;{\rm fm}^{-1}$ are the two roots of the 
quadratic equation 
\begin{equation}
 \beta^2-\frac{2}{r_0}\beta-\alpha^2=0.
 \label{eq:eq8}
\end{equation}  
The phase shifts using this approximation are positive at all energies, 
and this is reflected in Eq.\ (\ref{eq:eq7}) where $\lambda v^2(k)$ 
is attractive for all $k$.  From Fig.\ \ref{fig:fig1} we see that 
below $k_F=0.5\;{\rm fm}^{-1}$ the energy gap can with reasonable 
accuracy be calculated with the interaction obtained directly from 
the effective range approximation.  
One can therefore say that 
at densities below $k_F=0.5\;{\rm fm}^{-1}$, and at the crudest level 
of sophistication in many-body theory,  the superfluid properties 
of neutron matter are determined by just two parameters, namely 
the free-space scattering length and effective range. At such densities,
more complicated many-body terms are also less important.
Also interesting is the fact that the phase shifts predict the position 
of the first zero of $\Delta(k)$ in momentum space, since we see from 
Eq.\ (\ref{eq:eq4}) that $\Delta(k)=\Delta_F v(k)=0$ first for $\delta(k)=0$, 
which occurs at $E_{\rm lab}\approx 248$ MeV (pp scattering) 
corresponding to $k\approx 
1.73\;{\rm fm}^{-1}$.  This is in good agreement with the results of 
Khodel et al.\ \cite{khodel96}.  In Ref.\ \cite{khodel96} it is 
also shown that this first zero of the gap function determines the 
Fermi momentum at which $\Delta_F=0$.  Our results therefore indicate 
that this Fermi momentum is in fact given by the energy at which 
the $^1S_0$ phase shifts become negative. 

The calculation of the $^1S_0$ gap in symmetric nuclear matter is  
closely related to the one for neutron matter.  In fact, with 
charge-independent forces, like the older Bonn potentials, and 
free single--particle energies one would, of course, obtain 
exactly the same results.  However, the new potentials 
on the market are charge-dependent, in order to achieve high quality fits  
to both np and pp scattering data, and therefore we must solve three 
coupled gap equations for neutron-neutron (nn), proton-proton (pp), 
and neutron-proton (np) pairing \cite{good79}:
\begin{equation}
 \Delta_i(k)=-\frac{1}{\pi}\int_0^{\infty}dk'k'^2 V_i(k,k')
 \frac{\Delta_i(k')}{E(k')},
\label{eq:eq9}
\end{equation}
where $i$=nn, pp and np, and the quasiparticle energy is still given by 
$E(k)=\sqrt{(\epsilon(k)-\epsilon(k_F))^2+\Delta(k)^2}$, but the energy 
gap is now given by 
\begin{equation}
 \Delta(k)^2=\Delta_{\rm nn}(k)^2+\Delta_{\rm pp}(k)^2+\Delta_{\rm np}(k)^2.
\label{eq:eq10}
\end{equation}
Thus the equations are coupled through their common energy denominator.  
The $^1S_0$ pp and nn interactions are very nearly identical, so  
the set of equations above can be reduced to two: one for the nn 
(or pp) gap and one for 
the np gap.  Solving these equations, both with the CD-Bonn potential and 
with the phase shift approximations we get the results shown in Fig.\ 
\ref{fig:fig3}.  For comparison we have in the same figure plotted the 
results for pure neutron matter with the CD-Bonn potential (dashed line). 
From the figure it is clear that the phase shift approximation works well 
also in this case, and that the gap in symmetric matter is not very different 
from the gap in neutron matter.   As could be expected, the results 
are very close to those obtained earlier with charge-independent 
interactions \cite{baldo90,chen93,tak93,elg96}.

In summary, we have shown that in infinite neutron and nuclear matter, 
owing to the near rank-one separability of the NN interaction in 
the $^1S_0$ partial wave,  
we are able to compute the $^1S_0$ pairing gap directly from the NN 
phase shifts. This explains also why all NN potentials which fit 
the scattering data
result in almost identical $^1S_0$ pairing gaps.  
This is the main result of this work.
However, it should be mentioned that this result    
is not likely to survive in a more refined calculation, for instance 
if one includes density and spin-density fluctuations in the 
effective pairing interaction like in e.g., Refs.\ \cite{wam93,schulze96}.  
Other partial waves will then be involved, and the simple arguments 
employed here will no longer apply.  Our reasoning here
applies also only
to a partial wave where the $T$-matrix (almost) has a pole, and we have
neglected the fact that the phase shifts become negative at higher
energies.  
As a curiosity, we have found that at Fermi momenta below
$0.5$ fm$^{-1}$ the pairing gap is even determined by two parameters 
only, the effective range and the scattering length.
Also, we have pointed out that since the new NN interactions are charge 
dependent, one has to consider three coupled gap equations for 
$^1S_0$ pairing in nuclear matter.  The final result though is very 
nearly the same has what one obtains with charge-independent interactions.

We are much indebted to  J.\ W.\ Clark, and E.\ Osnes for many 
valuable comments and discussions.

\end{multicols}

\clearpage

\begin{figure}
    \input{fig2.tex}
    \caption{$^1S_0$ energy gap in neutron matter calculated with  
             the CD-Bonn potential compared with the direct calculation 
             from $^1S_0$ phase shifts.}
    \label{fig:fig2}
\end{figure}
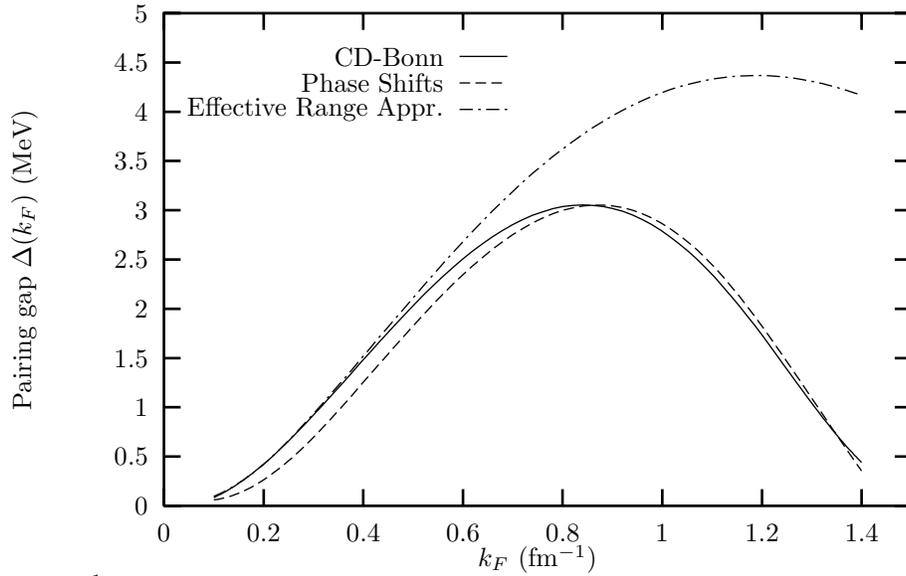 

\begin{figure}
    \input{fig3.tex}
	\caption{$^1S_0$ energy gap in nuclear matter calculated with  
                 the CD-Bonn potential  
                 compared with the direct calculation from the 
                 $^1S_0$ np and pp phase shifts.  Also shown are the results 
     for neutron matter with the CD-Bonn potential.}
    \label{fig:fig3}
\end{figure}
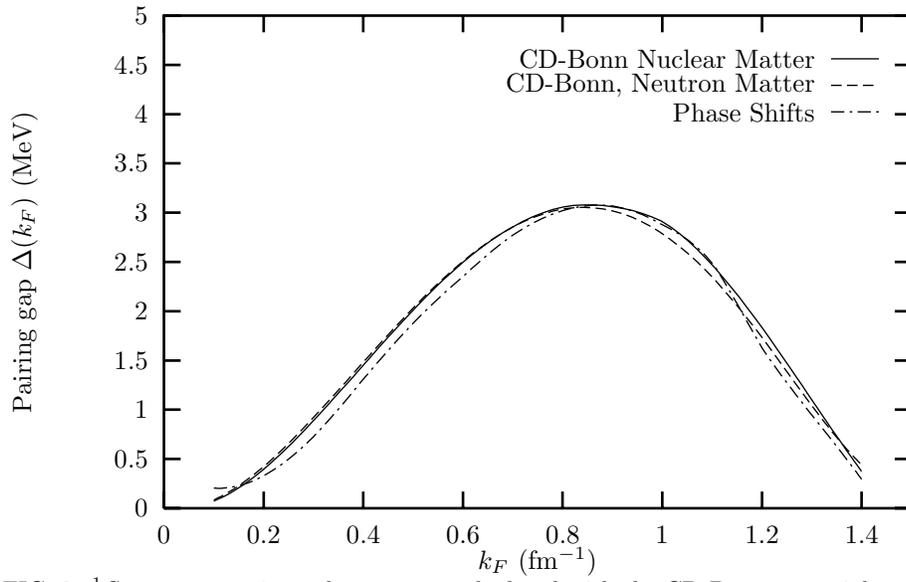	

\end{document}

%% file: fig1.tex
% GNUPLOT: LaTeX picture with Postscript
\setlength{\unitlength}{0.1bp}
\special{!
%!PS-Adobe-2.0
%%Creator: gnuplot
%%DocumentFonts: Helvetica
%%BoundingBox: 50 50 770 554
%%Pages: (atend)
%%EndComments
/gnudict 40 dict def
gnudict begin
/Color false def
/Solid false def
/gnulinewidth 5.000 def
/vshift -33 def
/dl {10 mul} def
/hpt 31.5 def
/vpt 31.5 def
/M {moveto} bind def
/L {lineto} bind def
/R {rmoveto} bind def
/V {rlineto} bind def
/vpt2 vpt 2 mul def
/hpt2 hpt 2 mul def
/Lshow { currentpoint stroke M
  0 vshift R show } def
/Rshow { currentpoint stroke M
  dup stringwidth pop neg vshift R show } def
/Cshow { currentpoint stroke M
  dup stringwidth pop -2 div vshift R show } def
/DL { Color {setrgbcolor Solid {pop []} if 0 setdash }
 {pop pop pop Solid {pop []} if 0 setdash} ifelse } def
/BL { stroke gnulinewidth 2 mul setlinewidth } def
/AL { stroke gnulinewidth 2 div setlinewidth } def
/PL { stroke gnulinewidth setlinewidth } def
/LTb { BL [] 0 0 0 DL } def
/LTa { AL [1 dl 2 dl] 0 setdash 0 0 0 setrgbcolor } def
/LT0 { PL [] 0 1 0 DL } def
/LT1 { PL [4 dl 2 dl] 0 0 1 DL } def
/LT2 { PL [2 dl 3 dl] 1 0 0 DL } def
/LT3 { PL [1 dl 1.5 dl] 1 0 1 DL } def
/LT4 { PL [5 dl 2 dl 1 dl 2 dl] 0 1 1 DL } def
/LT5 { PL [4 dl 3 dl 1 dl 3 dl] 1 1 0 DL } def
/LT6 { PL [2 dl 2 dl 2 dl 4 dl] 0 0 0 DL } def
/LT7 { PL [2 dl 2 dl 2 dl 2 dl 2 dl 4 dl] 1 0.3 0 DL } def
/LT8 { PL [2 dl 2 dl 2 dl 2 dl 2 dl 2 dl 2 dl 4 dl] 0.5 0.5 0.5 DL } def
/P { stroke [] 0 setdash
  currentlinewidth 2 div sub M
  0 currentlinewidth V stroke } def
/D { stroke [] 0 setdash 2 copy vpt add M
  hpt neg vpt neg V hpt vpt neg V
  hpt vpt V hpt neg vpt V closepath stroke
  P } def
/A { stroke [] 0 setdash vpt sub M 0 vpt2 V
  currentpoint stroke M
  hpt neg vpt neg R hpt2 0 V stroke
  } def
/B { stroke [] 0 setdash 2 copy exch hpt sub exch vpt add M
  0 vpt2 neg V hpt2 0 V 0 vpt2 V
  hpt2 neg 0 V closepath stroke
  P } def
/C { stroke [] 0 setdash exch hpt sub exch vpt add M
  hpt2 vpt2 neg V currentpoint stroke M
  hpt2 neg 0 R hpt2 vpt2 V stroke } def
/T { stroke [] 0 setdash 2 copy vpt 1.12 mul add M
  hpt neg vpt -1.62 mul V
  hpt 2 mul 0 V
  hpt neg vpt 1.62 mul V closepath stroke
  P  } def
/S { 2 copy A C} def
end
}
\begin{picture}(3600,2160)(0,0)
\special{"
gnudict begin
gsave
50 50 translate
0.100 0.100 scale
0 setgray
/Helvetica findfont 100 scalefont setfont
newpath
-500.000000 -500.000000 translate
LTa
600 251 M
2817 0 V
600 251 M
0 1858 V
LTb
600 251 M
63 0 V
2754 0 R
-63 0 V
600 437 M
63 0 V
2754 0 R
-63 0 V
600 623 M
63 0 V
2754 0 R
-63 0 V
600 808 M
63 0 V
2754 0 R
-63 0 V
600 994 M
63 0 V
2754 0 R
-63 0 V
600 1180 M
63 0 V
2754 0 R
-63 0 V
600 1366 M
63 0 V
2754 0 R
-63 0 V
600 1552 M
63 0 V
2754 0 R
-63 0 V
600 1737 M
63 0 V
2754 0 R
-63 0 V
600 1923 M
63 0 V
2754 0 R
-63 0 V
600 2109 M
63 0 V
2754 0 R
-63 0 V
600 251 M
0 63 V
0 1795 R
0 -63 V
976 251 M
0 63 V
0 1795 R
0 -63 V
1351 251 M
0 63 V
0 1795 R
0 -63 V
1727 251 M
0 63 V
0 1795 R
0 -63 V
2102 251 M
0 63 V
0 1795 R
0 -63 V
2478 251 M
0 63 V
0 1795 R
0 -63 V
2854 251 M
0 63 V
0 1795 R
0 -63 V
3229 251 M
0 63 V
0 1795 R
0 -63 V
600 251 M
2817 0 V
0 1858 V
-2817 0 V
600 251 L
LT0
3114 1946 M
180 0 V
788 282 M
2 1 V
2 1 V
4 2 V
5 3 V
6 3 V
8 3 V
8 5 V
10 5 V
11 7 V
12 7 V
13 8 V
14 10 V
16 11 V
16 12 V
18 13 V
18 15 V
20 16 V
21 18 V
21 19 V
23 22 V
23 22 V
25 25 V
25 26 V
26 27 V
27 29 V
28 30 V
29 32 V
29 32 V
31 34 V
31 35 V
31 36 V
33 36 V
32 36 V
34 37 V
34 37 V
35 37 V
35 37 V
35 36 V
36 35 V
36 34 V
37 33 V
37 32 V
37 30 V
38 28 V
37 26 V
38 24 V
38 22 V
38 19 V
38 17 V
39 13 V
38 12 V
38 8 V
38 4 V
38 2 V
37 -2 V
38 -4 V
37 -8 V
37 -10 V
37 -14 V
36 -17 V
36 -19 V
35 -22 V
35 -25 V
35 -27 V
34 -28 V
34 -31 V
32 -32 V
33 -34 V
31 -36 V
31 -36 V
31 -37 V
29 -37 V
29 -36 V
28 -37 V
27 -37 V
26 -36 V
25 -35 V
25 -34 V
23 -32 V
23 -31 V
21 -29 V
21 -28 V
20 -26 V
18 -24 V
18 -22 V
16 -21 V
16 -19 V
14 -17 V
13 -15 V
12 -14 V
11 -12 V
10 -10 V
8 -10 V
8 -7 V
6 -7 V
5 -5 V
4 -4 V
2 -3 V
2 -1 V
LT1
3114 1846 M
180 0 V
788 281 M
2 1 V
2 1 V
4 2 V
5 2 V
6 3 V
8 4 V
8 4 V
10 5 V
11 6 V
12 8 V
13 8 V
14 9 V
16 10 V
16 12 V
18 13 V
18 15 V
20 16 V
21 18 V
21 19 V
23 21 V
23 22 V
25 24 V
25 26 V
26 27 V
27 28 V
28 30 V
29 31 V
29 32 V
31 34 V
31 34 V
31 35 V
33 36 V
32 36 V
34 37 V
34 36 V
35 36 V
35 36 V
35 36 V
36 35 V
36 34 V
37 32 V
37 31 V
37 30 V
38 28 V
37 25 V
38 24 V
38 21 V
38 19 V
38 16 V
39 13 V
38 10 V
38 7 V
38 5 V
38 1 V
37 -2 V
38 -5 V
37 -8 V
37 -12 V
37 -14 V
36 -17 V
36 -20 V
35 -22 V
35 -25 V
35 -27 V
34 -30 V
34 -31 V
32 -32 V
33 -34 V
31 -36 V
31 -36 V
31 -36 V
29 -37 V
29 -37 V
28 -36 V
27 -37 V
26 -35 V
25 -35 V
25 -33 V
23 -32 V
23 -30 V
21 -29 V
21 -27 V
20 -25 V
18 -24 V
18 -21 V
16 -20 V
16 -18 V
14 -17 V
13 -14 V
12 -14 V
11 -11 V
10 -10 V
8 -9 V
8 -7 V
6 -6 V
5 -5 V
4 -4 V
2 -2 V
2 -2 V
LT2
3114 1746 M
180 0 V
788 282 M
2 1 V
2 1 V
4 2 V
5 2 V
6 3 V
8 4 V
8 5 V
10 5 V
11 6 V
12 8 V
13 8 V
14 10 V
16 10 V
16 12 V
18 14 V
18 14 V
20 17 V
21 18 V
21 19 V
23 21 V
23 23 V
25 24 V
25 26 V
26 28 V
27 28 V
28 31 V
29 31 V
29 33 V
31 34 V
31 35 V
31 36 V
33 36 V
32 36 V
34 37 V
34 37 V
35 37 V
35 37 V
35 36 V
36 35 V
36 35 V
37 33 V
37 32 V
37 30 V
38 29 V
37 26 V
38 25 V
38 22 V
38 19 V
38 17 V
39 14 V
38 12 V
38 8 V
38 5 V
38 2 V
37 -1 V
38 -4 V
37 -8 V
37 -10 V
37 -14 V
36 -16 V
36 -19 V
35 -22 V
35 -25 V
35 -26 V
34 -29 V
34 -31 V
32 -32 V
33 -34 V
31 -35 V
31 -37 V
31 -36 V
29 -37 V
29 -37 V
28 -37 V
27 -37 V
26 -36 V
25 -36 V
25 -34 V
23 -32 V
23 -32 V
21 -29 V
21 -28 V
20 -26 V
18 -25 V
18 -22 V
16 -21 V
16 -19 V
14 -18 V
13 -15 V
12 -14 V
11 -12 V
10 -11 V
8 -10 V
8 -7 V
6 -7 V
5 -5 V
4 -4 V
2 -3 V
2 -1 V
stroke
grestore
end
showpage
}
\put(3054,1746){\makebox(0,0)[r]{Nijmegen II}}
\put(3054,1846){\makebox(0,0)[r]{Nijmegen I}}
\put(3054,1946){\makebox(0,0)[r]{CD-Bonn}}
\put(2008,51){\makebox(0,0){$k_F$ (fm$^{-1}$)}}
\put(100,1180){%
\special{ps: gsave currentpoint currentpoint translate
270 rotate neg exch neg exch translate}%
\makebox(0,0)[b]{\shortstack{Pairing gap $\Delta (k_F)$ (MeV)}}%
\special{ps: currentpoint grestore moveto}%
}
\put(3229,151){\makebox(0,0){1.4}}
\put(2854,151){\makebox(0,0){1.2}}
\put(2478,151){\makebox(0,0){1}}
\put(2102,151){\makebox(0,0){0.8}}
\put(1727,151){\makebox(0,0){0.6}}
\put(1351,151){\makebox(0,0){0.4}}
\put(976,151){\makebox(0,0){0.2}}
\put(600,151){\makebox(0,0){0}}
\put(540,2109){\makebox(0,0)[r]{5}}
\put(540,1923){\makebox(0,0)[r]{4.5}}
\put(540,1737){\makebox(0,0)[r]{4}}
\put(540,1552){\makebox(0,0)[r]{3.5}}
\put(540,1366){\makebox(0,0)[r]{3}}
\put(540,1180){\makebox(0,0)[r]{2.5}}
\put(540,994){\makebox(0,0)[r]{2}}
\put(540,808){\makebox(0,0)[r]{1.5}}
\put(540,623){\makebox(0,0)[r]{1}}
\put(540,437){\makebox(0,0)[r]{0.5}}
\put(540,251){\makebox(0,0)[r]{0}}
\end{picture}

%% file: fig2.tex
% GNUPLOT: LaTeX picture with Postscript
\setlength{\unitlength}{0.1bp}
\special{!
%!PS-Adobe-2.0
%%Creator: gnuplot
%%DocumentFonts: Helvetica
%%BoundingBox: 50 50 770 554
%%Pages: (atend)
%%EndComments
/gnudict 40 dict def
gnudict begin
/Color false def
/Solid false def
/gnulinewidth 5.000 def
/vshift -33 def
/dl {10 mul} def
/hpt 31.5 def
/vpt 31.5 def
/M {moveto} bind def
/L {lineto} bind def
/R {rmoveto} bind def
/V {rlineto} bind def
/vpt2 vpt 2 mul def
/hpt2 hpt 2 mul def
/Lshow { currentpoint stroke M
  0 vshift R show } def
/Rshow { currentpoint stroke M
  dup stringwidth pop neg vshift R show } def
/Cshow { currentpoint stroke M
  dup stringwidth pop -2 div vshift R show } def
/DL { Color {setrgbcolor Solid {pop []} if 0 setdash }
 {pop pop pop Solid {pop []} if 0 setdash} ifelse } def
/BL { stroke gnulinewidth 2 mul setlinewidth } def
/AL { stroke gnulinewidth 2 div setlinewidth } def
/PL { stroke gnulinewidth setlinewidth } def
/LTb { BL [] 0 0 0 DL } def
/LTa { AL [1 dl 2 dl] 0 setdash 0 0 0 setrgbcolor } def
/LT0 { PL [] 0 1 0 DL } def
/LT1 { PL [4 dl 2 dl] 0 0 1 DL } def
/LT2 { PL [2 dl 3 dl] 1 0 0 DL } def
/LT3 { PL [1 dl 1.5 dl] 1 0 1 DL } def
/LT4 { PL [5 dl 2 dl 1 dl 2 dl] 0 1 1 DL } def
/LT5 { PL [4 dl 3 dl 1 dl 3 dl] 1 1 0 DL } def
/LT6 { PL [2 dl 2 dl 2 dl 4 dl] 0 0 0 DL } def
/LT7 { PL [2 dl 2 dl 2 dl 2 dl 2 dl 4 dl] 1 0.3 0 DL } def
/LT8 { PL [2 dl 2 dl 2 dl 2 dl 2 dl 2 dl 2 dl 4 dl] 0.5 0.5 0.5 DL } def
/P { stroke [] 0 setdash
  currentlinewidth 2 div sub M
  0 currentlinewidth V stroke } def
/D { stroke [] 0 setdash 2 copy vpt add M
  hpt neg vpt neg V hpt vpt neg V
  hpt vpt V hpt neg vpt V closepath stroke
  P } def
/A { stroke [] 0 setdash vpt sub M 0 vpt2 V
  currentpoint stroke M
  hpt neg vpt neg R hpt2 0 V stroke
  } def
/B { stroke [] 0 setdash 2 copy exch hpt sub exch vpt add M
  0 vpt2 neg V hpt2 0 V 0 vpt2 V
  hpt2 neg 0 V closepath stroke
  P } def
/C { stroke [] 0 setdash exch hpt sub exch vpt add M
  hpt2 vpt2 neg V currentpoint stroke M
  hpt2 neg 0 R hpt2 vpt2 V stroke } def
/T { stroke [] 0 setdash 2 copy vpt 1.12 mul add M
  hpt neg vpt -1.62 mul V
  hpt 2 mul 0 V
  hpt neg vpt 1.62 mul V closepath stroke
  P  } def
/S { 2 copy A C} def
end
}
\begin{picture}(3600,2160)(0,0)
\special{"
gnudict begin
gsave
50 50 translate
0.100 0.100 scale
0 setgray
/Helvetica findfont 100 scalefont setfont
newpath
-500.000000 -500.000000 translate
LTa
600 251 M
2817 0 V
600 251 M
0 1858 V
LTb
600 251 M
63 0 V
2754 0 R
-63 0 V
600 437 M
63 0 V
2754 0 R
-63 0 V
600 623 M
63 0 V
2754 0 R
-63 0 V
600 808 M
63 0 V
2754 0 R
-63 0 V
600 994 M
63 0 V
2754 0 R
-63 0 V
600 1180 M
63 0 V
2754 0 R
-63 0 V
600 1366 M
63 0 V
2754 0 R
-63 0 V
600 1552 M
63 0 V
2754 0 R
-63 0 V
600 1737 M
63 0 V
2754 0 R
-63 0 V
600 1923 M
63 0 V
2754 0 R
-63 0 V
600 2109 M
63 0 V
2754 0 R
-63 0 V
600 251 M
0 63 V
0 1795 R
0 -63 V
976 251 M
0 63 V
0 1795 R
0 -63 V
1351 251 M
0 63 V
0 1795 R
0 -63 V
1727 251 M
0 63 V
0 1795 R
0 -63 V
2102 251 M
0 63 V
0 1795 R
0 -63 V
2478 251 M
0 63 V
0 1795 R
0 -63 V
2854 251 M
0 63 V
0 1795 R
0 -63 V
3229 251 M
0 63 V
0 1795 R
0 -63 V
600 251 M
2817 0 V
0 1858 V
-2817 0 V
600 251 L
LT0
1714 1946 M
180 0 V
788 282 M
2 1 V
2 1 V
4 2 V
5 3 V
6 3 V
8 3 V
8 5 V
10 5 V
11 7 V
12 7 V
13 8 V
14 10 V
16 11 V
16 12 V
18 13 V
18 15 V
20 16 V
21 18 V
21 19 V
23 22 V
23 22 V
25 25 V
25 26 V
26 27 V
27 29 V
28 30 V
29 32 V
29 32 V
31 34 V
31 35 V
31 36 V
33 36 V
32 36 V
34 37 V
34 37 V
35 37 V
35 37 V
35 36 V
36 35 V
36 34 V
37 33 V
37 32 V
37 30 V
38 28 V
37 26 V
38 24 V
38 22 V
38 19 V
38 17 V
39 13 V
38 12 V
38 8 V
38 4 V
38 2 V
37 -2 V
38 -4 V
37 -8 V
37 -10 V
37 -14 V
36 -17 V
36 -19 V
35 -22 V
35 -25 V
35 -27 V
34 -28 V
34 -31 V
32 -32 V
33 -34 V
31 -36 V
31 -36 V
31 -37 V
29 -37 V
29 -36 V
28 -37 V
27 -37 V
26 -36 V
25 -35 V
25 -34 V
23 -32 V
23 -31 V
21 -29 V
21 -28 V
20 -26 V
18 -24 V
18 -22 V
16 -21 V
16 -19 V
14 -17 V
13 -15 V
12 -14 V
11 -12 V
10 -10 V
8 -10 V
8 -7 V
6 -7 V
5 -5 V
4 -4 V
2 -3 V
2 -1 V
LT1
1714 1846 M
180 0 V
788 274 M
2 0 V
2 0 V
4 1 V
5 0 V
6 1 V
8 2 V
8 2 V
10 2 V
11 3 V
12 4 V
13 5 V
14 5 V
16 7 V
16 8 V
18 9 V
18 11 V
20 12 V
21 14 V
21 16 V
23 17 V
23 19 V
25 21 V
25 23 V
26 25 V
27 27 V
28 29 V
29 31 V
29 32 V
31 34 V
31 35 V
31 37 V
33 37 V
32 37 V
34 38 V
34 39 V
35 39 V
35 39 V
35 38 V
36 39 V
36 37 V
37 37 V
37 35 V
37 34 V
38 32 V
37 31 V
38 28 V
38 27 V
38 23 V
38 22 V
39 19 V
38 16 V
38 13 V
38 9 V
38 7 V
37 3 V
38 0 V
37 -3 V
37 -6 V
37 -10 V
36 -13 V
36 -16 V
35 -19 V
35 -22 V
35 -25 V
34 -27 V
34 -29 V
32 -32 V
33 -33 V
31 -35 V
31 -37 V
31 -37 V
29 -38 V
29 -38 V
28 -38 V
27 -38 V
26 -37 V
25 -36 V
25 -36 V
23 -34 V
23 -33 V
21 -32 V
21 -31 V
20 -29 V
18 -27 V
18 -26 V
16 -25 V
16 -22 V
14 -21 V
13 -19 V
12 -18 V
11 -16 V
10 -14 V
8 -12 V
8 -11 V
6 -9 V
5 -7 V
4 -5 V
2 -4 V
2 -2 V
LT4
1714 1746 M
180 0 V
788 287 M
2 1 V
2 1 V
4 1 V
5 3 V
6 2 V
8 4 V
8 4 V
10 5 V
11 6 V
12 7 V
13 8 V
14 9 V
16 10 V
16 12 V
18 13 V
18 15 V
20 16 V
21 18 V
21 20 V
23 22 V
23 23 V
25 25 V
25 27 V
26 28 V
27 30 V
28 31 V
29 33 V
29 34 V
31 35 V
31 36 V
31 38 V
33 38 V
32 39 V
34 39 V
34 40 V
35 40 V
35 41 V
35 41 V
36 41 V
36 41 V
37 41 V
37 40 V
37 40 V
38 39 V
37 38 V
38 37 V
38 36 V
38 36 V
38 34 V
39 32 V
38 32 V
38 29 V
38 28 V
38 27 V
37 25 V
38 23 V
37 22 V
37 20 V
37 19 V
36 17 V
36 16 V
35 13 V
35 13 V
35 11 V
34 9 V
34 8 V
32 7 V
33 6 V
31 4 V
31 4 V
31 2 V
29 1 V
29 1 V
28 0 V
27 -1 V
26 -2 V
25 -2 V
25 -3 V
23 -3 V
23 -4 V
21 -4 V
21 -4 V
20 -4 V
18 -4 V
18 -5 V
16 -4 V
16 -4 V
14 -5 V
13 -4 V
12 -3 V
11 -4 V
10 -3 V
8 -3 V
8 -2 V
6 -3 V
5 -1 V
4 -2 V
2 -1 V
2 0 V
stroke
grestore
end
showpage
}
\put(1654,1746){\makebox(0,0)[r]{Effective Range Appr.}}
\put(1654,1846){\makebox(0,0)[r]{Phase Shifts}}
\put(1654,1946){\makebox(0,0)[r]{CD-Bonn}}
\put(2008,51){\makebox(0,0){$k_F$ (fm$^{-1}$)}}
\put(100,1180){%
\special{ps: gsave currentpoint currentpoint translate
270 rotate neg exch neg exch translate}%
\makebox(0,0)[b]{\shortstack{Pairing gap $\Delta (k_F)$ (MeV)}}%
\special{ps: currentpoint grestore moveto}%
}
\put(3229,151){\makebox(0,0){1.4}}
\put(2854,151){\makebox(0,0){1.2}}
\put(2478,151){\makebox(0,0){1}}
\put(2102,151){\makebox(0,0){0.8}}
\put(1727,151){\makebox(0,0){0.6}}
\put(1351,151){\makebox(0,0){0.4}}
\put(976,151){\makebox(0,0){0.2}}
\put(600,151){\makebox(0,0){0}}
\put(540,2109){\makebox(0,0)[r]{5}}
\put(540,1923){\makebox(0,0)[r]{4.5}}
\put(540,1737){\makebox(0,0)[r]{4}}
\put(540,1552){\makebox(0,0)[r]{3.5}}
\put(540,1366){\makebox(0,0)[r]{3}}
\put(540,1180){\makebox(0,0)[r]{2.5}}
\put(540,994){\makebox(0,0)[r]{2}}
\put(540,808){\makebox(0,0)[r]{1.5}}
\put(540,623){\makebox(0,0)[r]{1}}
\put(540,437){\makebox(0,0)[r]{0.5}}
\put(540,251){\makebox(0,0)[r]{0}}
\end{picture}

%% file: fig3.tex
% GNUPLOT: LaTeX picture with Postscript
\setlength{\unitlength}{0.1bp}
\special{!
%!PS-Adobe-2.0
%%Creator: gnuplot
%%DocumentFonts: Helvetica
%%BoundingBox: 50 50 770 554
%%Pages: (atend)
%%EndComments
/gnudict 40 dict def
gnudict begin
/Color false def
/Solid false def
/gnulinewidth 5.000 def
/vshift -33 def
/dl {10 mul} def
/hpt 31.5 def
/vpt 31.5 def
/M {moveto} bind def
/L {lineto} bind def
/R {rmoveto} bind def
/V {rlineto} bind def
/vpt2 vpt 2 mul def
/hpt2 hpt 2 mul def
/Lshow { currentpoint stroke M
  0 vshift R show } def
/Rshow { currentpoint stroke M
  dup stringwidth pop neg vshift R show } def
/Cshow { currentpoint stroke M
  dup stringwidth pop -2 div vshift R show } def
/DL { Color {setrgbcolor Solid {pop []} if 0 setdash }
 {pop pop pop Solid {pop []} if 0 setdash} ifelse } def
/BL { stroke gnulinewidth 2 mul setlinewidth } def
/AL { stroke gnulinewidth 2 div setlinewidth } def
/PL { stroke gnulinewidth setlinewidth } def
/LTb { BL [] 0 0 0 DL } def
/LTa { AL [1 dl 2 dl] 0 setdash 0 0 0 setrgbcolor } def
/LT0 { PL [] 0 1 0 DL } def
/LT1 { PL [4 dl 2 dl] 0 0 1 DL } def
/LT2 { PL [2 dl 3 dl] 1 0 0 DL } def
/LT3 { PL [1 dl 1.5 dl] 1 0 1 DL } def
/LT4 { PL [5 dl 2 dl 1 dl 2 dl] 0 1 1 DL } def
/LT5 { PL [4 dl 3 dl 1 dl 3 dl] 1 1 0 DL } def
/LT6 { PL [2 dl 2 dl 2 dl 4 dl] 0 0 0 DL } def
/LT7 { PL [2 dl 2 dl 2 dl 2 dl 2 dl 4 dl] 1 0.3 0 DL } def
/LT8 { PL [2 dl 2 dl 2 dl 2 dl 2 dl 2 dl 2 dl 4 dl] 0.5 0.5 0.5 DL } def
/P { stroke [] 0 setdash
  currentlinewidth 2 div sub M
  0 currentlinewidth V stroke } def
/D { stroke [] 0 setdash 2 copy vpt add M
  hpt neg vpt neg V hpt vpt neg V
  hpt vpt V hpt neg vpt V closepath stroke
  P } def
/A { stroke [] 0 setdash vpt sub M 0 vpt2 V
  currentpoint stroke M
  hpt neg vpt neg R hpt2 0 V stroke
  } def
/B { stroke [] 0 setdash 2 copy exch hpt sub exch vpt add M
  0 vpt2 neg V hpt2 0 V 0 vpt2 V
  hpt2 neg 0 V closepath stroke
  P } def
/C { stroke [] 0 setdash exch hpt sub exch vpt add M
  hpt2 vpt2 neg V currentpoint stroke M
  hpt2 neg 0 R hpt2 vpt2 V stroke } def
/T { stroke [] 0 setdash 2 copy vpt 1.12 mul add M
  hpt neg vpt -1.62 mul V
  hpt 2 mul 0 V
  hpt neg vpt 1.62 mul V closepath stroke
  P  } def
/S { 2 copy A C} def
end
}
\begin{picture}(3600,2160)(0,0)
\special{"
gnudict begin
gsave
50 50 translate
0.100 0.100 scale
0 setgray
/Helvetica findfont 100 scalefont setfont
newpath
-500.000000 -500.000000 translate
LTa
600 251 M
2817 0 V
600 251 M
0 1858 V
LTb
600 251 M
63 0 V
2754 0 R
-63 0 V
600 437 M
63 0 V
2754 0 R
-63 0 V
600 623 M
63 0 V
2754 0 R
-63 0 V
600 808 M
63 0 V
2754 0 R
-63 0 V
600 994 M
63 0 V
2754 0 R
-63 0 V
600 1180 M
63 0 V
2754 0 R
-63 0 V
600 1366 M
63 0 V
2754 0 R
-63 0 V
600 1552 M
63 0 V
2754 0 R
-63 0 V
600 1737 M
63 0 V
2754 0 R
-63 0 V
600 1923 M
63 0 V
2754 0 R
-63 0 V
600 2109 M
63 0 V
2754 0 R
-63 0 V
600 251 M
0 63 V
0 1795 R
0 -63 V
976 251 M
0 63 V
0 1795 R
0 -63 V
1351 251 M
0 63 V
0 1795 R
0 -63 V
1727 251 M
0 63 V
0 1795 R
0 -63 V
2102 251 M
0 63 V
0 1795 R
0 -63 V
2478 251 M
0 63 V
0 1795 R
0 -63 V
2854 251 M
0 63 V
0 1795 R
0 -63 V
3229 251 M
0 63 V
0 1795 R
0 -63 V
600 251 M
2817 0 V
0 1858 V
-2817 0 V
600 251 L
LT0
3114 1946 M
180 0 V
788 278 M
2 1 V
2 1 V
4 2 V
5 2 V
6 3 V
8 3 V
8 4 V
10 5 V
11 6 V
12 7 V
13 8 V
14 9 V
16 10 V
16 12 V
18 13 V
18 14 V
20 16 V
21 18 V
21 19 V
23 21 V
23 22 V
25 24 V
25 26 V
26 27 V
27 29 V
28 30 V
29 32 V
29 33 V
31 34 V
31 34 V
31 36 V
33 37 V
32 37 V
34 38 V
34 37 V
35 38 V
35 37 V
35 37 V
36 36 V
36 35 V
37 33 V
37 33 V
37 31 V
38 29 V
37 27 V
38 25 V
38 22 V
38 21 V
38 18 V
39 15 V
38 13 V
38 9 V
38 6 V
38 2 V
37 0 V
38 -2 V
37 -5 V
37 -6 V
37 -8 V
36 -11 V
36 -13 V
35 -15 V
35 -22 V
35 -26 V
34 -29 V
34 -32 V
32 -33 V
33 -35 V
31 -37 V
31 -37 V
31 -38 V
29 -39 V
29 -38 V
28 -38 V
27 -38 V
26 -38 V
25 -36 V
25 -36 V
23 -34 V
23 -33 V
21 -32 V
21 -31 V
20 -29 V
18 -27 V
18 -26 V
16 -24 V
16 -22 V
14 -20 V
13 -19 V
12 -18 V
11 -15 V
10 -14 V
8 -12 V
8 -10 V
6 -9 V
5 -7 V
4 -5 V
2 -4 V
2 -2 V
LT1
3114 1846 M
180 0 V
788 282 M
2 1 V
2 1 V
4 2 V
5 3 V
6 3 V
8 3 V
8 5 V
10 5 V
11 7 V
12 7 V
13 8 V
14 10 V
16 11 V
16 12 V
18 13 V
18 15 V
20 16 V
21 18 V
21 19 V
23 22 V
23 22 V
25 25 V
25 26 V
26 27 V
27 29 V
28 30 V
29 32 V
29 32 V
31 34 V
31 35 V
31 36 V
33 36 V
32 36 V
34 37 V
34 37 V
35 37 V
35 37 V
35 36 V
36 35 V
36 34 V
37 33 V
37 32 V
37 30 V
38 28 V
37 26 V
38 24 V
38 22 V
38 19 V
38 17 V
39 13 V
38 12 V
38 8 V
38 4 V
38 2 V
37 -2 V
38 -4 V
37 -8 V
37 -10 V
37 -14 V
36 -17 V
36 -19 V
35 -22 V
35 -25 V
35 -27 V
34 -28 V
34 -31 V
32 -32 V
33 -34 V
31 -36 V
31 -36 V
31 -37 V
29 -37 V
29 -36 V
28 -37 V
27 -37 V
26 -36 V
25 -35 V
25 -34 V
23 -32 V
23 -31 V
21 -29 V
21 -28 V
20 -26 V
18 -24 V
18 -22 V
16 -21 V
16 -19 V
14 -17 V
13 -15 V
12 -14 V
11 -12 V
10 -10 V
8 -10 V
8 -7 V
6 -7 V
5 -5 V
4 -4 V
2 -3 V
2 -1 V
LT4
3114 1746 M
180 0 V
788 327 M
2 0 V
2 0 V
4 0 V
5 -1 V
6 0 V
8 0 V
8 1 V
10 1 V
11 1 V
12 2 V
13 2 V
14 4 V
16 4 V
16 5 V
18 7 V
18 8 V
20 10 V
21 12 V
21 13 V
23 15 V
23 17 V
25 20 V
25 21 V
26 24 V
27 26 V
28 30 V
29 32 V
29 33 V
31 36 V
31 37 V
31 39 V
33 38 V
32 38 V
34 38 V
34 39 V
35 38 V
35 37 V
35 35 V
36 35 V
36 34 V
37 33 V
37 32 V
37 34 V
38 33 V
37 32 V
38 30 V
38 28 V
38 24 V
38 22 V
39 19 V
38 16 V
38 13 V
38 10 V
38 6 V
37 4 V
38 0 V
37 -3 V
37 -8 V
37 -11 V
36 -13 V
36 -17 V
35 -19 V
35 -18 V
35 -21 V
34 -23 V
34 -28 V
32 -31 V
33 -39 V
31 -50 V
31 -53 V
31 -53 V
29 -53 V
29 -51 V
28 -49 V
27 -40 V
26 -36 V
25 -35 V
25 -32 V
23 -31 V
23 -29 V
21 -28 V
21 -26 V
20 -25 V
18 -23 V
18 -22 V
16 -21 V
16 -20 V
14 -18 V
13 -17 V
12 -16 V
11 -14 V
10 -13 V
8 -11 V
8 -10 V
6 -9 V
5 -7 V
4 -5 V
2 -3 V
2 -3 V
stroke
grestore
end
showpage
}
\put(3054,1746){\makebox(0,0)[r]{Phase Shifts}}
\put(3054,1846){\makebox(0,0)[r]{CD-Bonn, Neutron Matter}}
\put(3054,1946){\makebox(0,0)[r]{CD-Bonn Nuclear Matter}}
\put(2008,51){\makebox(0,0){$k_F$ (fm$^{-1}$)}}
\put(100,1180){%
\special{ps: gsave currentpoint currentpoint translate
270 rotate neg exch neg exch translate}%
\makebox(0,0)[b]{\shortstack{Pairing gap $\Delta (k_F)$ (MeV)}}%
\special{ps: currentpoint grestore moveto}%
}
\put(3229,151){\makebox(0,0){1.4}}
\put(2854,151){\makebox(0,0){1.2}}
\put(2478,151){\makebox(0,0){1}}
\put(2102,151){\makebox(0,0){0.8}}
\put(1727,151){\makebox(0,0){0.6}}
\put(1351,151){\makebox(0,0){0.4}}
\put(976,151){\makebox(0,0){0.2}}
\put(600,151){\makebox(0,0){0}}
\put(540,2109){\makebox(0,0)[r]{5}}
\put(540,1923){\makebox(0,0)[r]{4.5}}
\put(540,1737){\makebox(0,0)[r]{4}}
\put(540,1552){\makebox(0,0)[r]{3.5}}
\put(540,1366){\makebox(0,0)[r]{3}}
\put(540,1180){\makebox(0,0)[r]{2.5}}
\put(540,994){\makebox(0,0)[r]{2}}
\put(540,808){\makebox(0,0)[r]{1.5}}
\put(540,623){\makebox(0,0)[r]{1}}
\put(540,437){\makebox(0,0)[r]{0.5}}
\put(540,251){\makebox(0,0)[r]{0}}
\end{picture}

%% file: paper.bbl
\begin{references}
\bibitem{petra95} C.\ J.\ Pethick and D.\ G.\ Ravenhall, Annu.\ Rev.\ 
Nucl.\ Part.\ Sci.\ {\bf 45}, 429 (1995).
\bibitem{mulshe93} A.\ C.\ M\"{u}ller and B.\ M.\ Sherril, Annu.\ Rev.\ Nucl.\ 
Part.\ Sci.\ {\bf 43}, 529 (1993).
\bibitem{riis94} K.\ Riisager, Rev.\ Mod.\ Phys.\ {\bf 66}, 1105 (1994).
\bibitem{baldo90} M.\ Baldo, J.\ Cugnon, A.\ Lejeune, and U.\ Lombardo, 
Nucl.\ Phys.\ A {\bf 515}, 409 (1990).
\bibitem{chen93} J.\ M.\ C.\ Chen, J.\ W.\ Clark, R.\ D.\ Dav\'{e}, and 
V.\ V.\ Khodel, Nucl.\ Phys.\ A {\bf 555}, 59 (1993).
\bibitem{tak93} T.\ Takatsuka and R.\ Tamagaki, Prog.\ Theor.\ Phys.\ 
Suppl.\ {\bf 112}, 27 (1993).
\bibitem{elg96} \O.\ Elgar\o y, L.\ Engvik, M.\ Hjorth-Jensen, and E.\ Osnes, 
Nucl.\ Phys.\ A {\bf 604}, 466 (1996).
\bibitem{khodel96} V.\ A.\ Khodel, V.\ V.\ Khodel, and J.\ W.\ Clark, 
Nucl.\ Phys.\ A {\bf 598}, 390 (1996).
\bibitem{ring90} H.\ Kucharek and P.\ Ring, Z.\ Phys. {\bf 339}, 23 (1990).
\bibitem{guim96} F.\ B.\ Guimar\~{a}es, B.\ V.\ Carlson, and T.\ Frederico, 
Phys.\ Rev.\ C {\bf 54}, 2385 (1996).
\bibitem{matera97} F.\ Matera, G.\ Fabbri, and A.\ Dellafiore, 
Phys.\ Rev.\ C {\bf 56}, 228 (1997).
\bibitem{clark76} J.\ W.\ Clark, C.\ G.\ K\"{a}llman, C.\-H.\ Yang, and 
D.\ A.\ Chakkalakal, Phys.\ Lett.\ B {\bf 61}, 331 (1976).
\bibitem{wam93} J.\ Wambach, T.\ L.\ Ainsworth, and D.\ Pines, Nucl.\ Phys.\ 
A {\bf 555}, 128 (1993).
\bibitem{schulze96} H.\-J.\ Schulze, J.\ Cugnon, A.\ Lejeune, M.\ Baldo, 
and U.\ Lombardo, Phys.\ Lett.\ B {\bf 375}, 1 (1996).
%\bibitem{carlson97} B.\ V.\ Carlson, T.\ Frederico, 
%and F.\ B.\ Guimar\~{a}es, nucl-th/9706071.
\bibitem{es60} V.\ J.\ Emery and A.\ M.\ Sessler,  Phys.\ Rev.\ {\bf 119},
248 (1960) and references therein.
\bibitem{hoffberg70} M.\ Hoffberg, A.\ E.\ Glassgold, R.\ W.\ Richardson, 
and M.\ Ruderman, Phys.\ Rev.\ Lett.\ {\bf 24}, 775 (1970).
\bibitem{mach96} R.\ Machleidt, F.\ Sammarruca, and Y.\ Song, Phys.\ Rev.\ C 
{\bf 53}, 1483 (1996).
\bibitem{nijm94} V.\ G.\ J.\ Stoks, R.\ A.\ M.\ Klomp, C.\ P.\ F.\ Terheggen, 
and J.\ J.\ de Swart, Phys.\ Rev.\ C {\bf 49}, 2950 (1994);
NN-Online facility, (URL: http://NN-OnLine.sci.kun.nl).
\bibitem{brown76} G.\ E.\ Brown and A.\ D.\ Jackson, {\it The Nucleon-Nucleon 
Interaction} (North-Holland, Amsterdam, 1976).
\bibitem{chadan92} K.\ Chadan and P.\ C.\ Sabatier, {\it Inverse Problems in 
Quantum Scattering Theory}, 2nd ed.\ (Springer, New York, 1992).
\bibitem{kk97} N.\ H.\ Kwong and H.\ S.\ K\"ohler, Phys.\ Rev.\ C {\bf 55},
1650 (1997).
\bibitem{nijm93} V.\ G.\ J.\ Stoks, R.\ A.\ M.\ Klomp, 
M.\ G.\ M.\ Rentmeester, and 
J.\ J.\ de Swart, Phys.\ Rev.\ C {\bf 48}, 792 (1993). 
\bibitem{davies91} K.\ T.\ R.\ Davies, G.\ D.\ White, and R.\ W.\ Davies, 
Nucl.\ Phys.\ A {\bf 524}, 743 (1991).
\bibitem{good79} A.\ L.\ Goodman, Adv.\ Nucl.\ Phys.\ {\bf 11}, 263 (1979).
\end{references}
